\documentclass[aps,prl,amsmath,amssymb,amsfonts,nofootinbib,long]{revtex4}
\usepackage{epsfig,latexsym,bm,makeidx}

\newcommand\Mpc{\mbox{Mpc}}

\newcommand\yr{\mbox{yr}}

\newcommand\as{\mbox{as}}
\newcommand{\bs}{\!\!\!\!\!}
\newcommand{\BS}{\!\!\!\!\!\!\!\!\!\!\!\!\!\!\!}
\newcommand\km{\mbox{km}}
\newcommand\s{\mbox{sec}}

\begin{document}

\title{COSMIC PARALLAX IN ELLIPSOIDAL UNIVERSE}

\author{L. Campanelli$^{1}$}
\email{leonardo.campanelli@ba.infn.it}

\author{P. Cea$^{1,2}$}
\email{paolo.cea@ba.infn.it}

\author{G. L. Fogli$^{1,2}$}
\email{gianluigi.fogli@ba.infn.it}

\author{L. Tedesco$^{1,2}$}
\email{luigi.tedesco@ba.infn.it}

\affiliation{$^1$Dipartimento di Fisica, Universit\`{a} di Bari, I-70126 Bari, Italy}
\affiliation{$^2$INFN - Sezione di Bari, I-70126 Bari, Italy}

\date{March, 2011}


\begin{abstract}
\noindent
{\bf Abstract.} The detection of a time variation of the angle between two distant sources would reveal an
anisotropic expansion of the Universe. We study this effect of {\it cosmic parallax} within
the {\it ellipsoidal universe} model, namely a particular homogeneous anisotropic cosmological
model of Bianchi type I, whose attractive feature is the potentiality to account for the
observed lack of power of the large-scale cosmic microwave background anisotropy.
The preferred direction in the sky, singled out by the axis of symmetry inherent to planar
symmetry of ellipsoidal universe, could in principle be constrained by future cosmic parallax
data. However, that will be a real possibility if and when the experimental accuracy will be
enhanced at least by two orders of magnitude.
\\
\\
\end{abstract}


\maketitle


\section{\normalsize{I. Introduction}}
\renewcommand{\thesection}{\arabic{section}}

Observational cosmology is entering a precision era thanks to, just to cite a few, the high
precision data collected by the Wilkinson Microwave Anisotropy Probe (WMAP) satellite~\cite{WMAP7a},
to the recently-launched PLANCK mission~\cite{PLANCK}, and to planned space projects such as
the Gaia astrometric mission~\cite{Gaia}.

Therefore, one may wonder if perhaps it is time to directly test the correctness of the
assumptions which are at the base of the by-now-accepted standard cosmological model,
the Lambda cold dark matter ($\Lambda$CDM) concordance model~\cite{Weinberg}, namely
homogeneity and isotropy of the large-scale structure of the Universe.

Indeed, some anomalous features in the seven-year WMAP data have already been interpreted
as a hint that our Universe could be not isotropic~\cite{prl} (see also
Refs.~\cite{prd,prd2,ADE,Berera,Barrow1,Barrow2,Koivisto-Mota,Koivisto-Mota2,Rodrigues,Akarsu1,Akarsu2,Akarsu3,Appleby,Battye,Watanabe}).

In particular, the analysis of inhomogeneities in the distribution of the excursion sets
in the cosmic microwave background (CMB) maps, suggests the existence of an anomalous
plane-mirroring symmetry on large angular scales~\cite{Gurzadyan1,Gurzadyan2}.

Moreover, the presence of a preferred direction in the sky, revealed by the alignment
of quadrupole and octupole modes of CMB anisotropy spectrum,
the so-called ``Axis of Evil'' (AE)~\cite{Land-Magueijo}, seems to indicate planar symmetry
in the geometry of the Universe on large cosmic scales.

Finally, the same CMB data show a suppression of power at large angular scales, i.e. a
quadrupole amplitude lower than that predicted by the $\Lambda$CDM model. This ``anomalous''
occurrence, refereed to as ``quadrupole problem'', has been widely studied in the
literature~\cite{Quadrupole1,Quadrupole2,Quadrupole3,Quadrupole4,Quadrupole5,Quadrupole6,Quadrupole7,Quadrupole8,Quadrupole9,Quadrupole10,Quadrupole11,Quadrupole12,Quadrupole13,Quadrupole14,Quadrupole15,Quadrupole16,Quadrupole17,Quadrupole18,Quadrupole19,Quadrupole20,Quadrupole21}.

A possible solution to that problem has been given in Ref.~\cite{prl} and rests on the idea
that the Universe is expanding anisotropically with respect to a certain direction in the sky.
The resulting cosmological model is essentially of Bianchi type I, where the free parameters
are fixed by requiring that the quadrupole generated by anisotropic expansion lowers, down to
the desired level, that caused by the standard inflationary mechanism. In the following, this
particular cosmological model will be referred to as ``ellipsoidal universe''.

It is only recently that the possibility has been put forward that an anisotropic expansion of
the Universe could be directly revealed by high-precision astrometric measurements. As shown in
the seminal paper by Quercellini, Quartin and Amendola~\cite{Quercellini}, space missions like
Gaia could measure tiny variations, over a period of a decade, of the relative angular position
of two bright sources in the sky, namely an effect of ``cosmic parallax''. If this will be the
case, far reaching consequences there will be on our vision of the structure of the Universe.

The aim of this paper is to analyze such a cosmic effect of parallax in the context of the
ellipsoidal universe.

The plan of the paper is then as follows. In section II, we point out all the essential properties
of a universe filled with a nonstandard fluid with anisotropic equation of state and we discuss
its relation to the ellipsoidal universe model. In section III, we discuss cosmic parallax
emphasizing its dependence on the direction of the axis of symmetry introduced by the ellipsoidal
universe model. In the Conclusions, we briefly comment on our results.


\section{\normalsize{II. Ellipsoidal Universe and Cosmic Shear}}
\renewcommand{\thesection}{\arabic{section}}

The ellipsoidal universe proposal~\cite{prl} rests on the assumption that
the large-scale geometry of the Universe is described by a Taub line
element~\cite{Taub}
\begin{equation}
\label{metric}
ds^2 = dt^2 - a^2(t) (dx^2 + dy^2) - b^2(t) \, dz^2,
\end{equation}
with two expansion parameters, $a$ and $b$, which we normalize as $a(t_0) = b(t_0) = 1$
at the present cosmic time $t_0$. The above metric is homogeneous but anisotropic and,
according to Bianchi classification, is of type I. The resulting universe possesses
(at large cosmological scales) planar symmetry, with axis of symmetry directed along the
$z$-axis.

In such a type of universe, a quadrupole term develops in the CMB radiation~\cite{prl}
which adds to that caused by the inflation-produced gravitational potential at the last
scattering surface. By a suitable orientation of the two terms,
it is possible to lower the overall quadrupole power to such an extent to
match the ``low'' value of the observed quadrupole to that predicted by
the standard cosmological model. Introducing the ``eccentricity''
\begin{equation}
\label{ecc-def} e =
    \left\{ \begin{array}{ll}
            \sqrt{1-(b/a)^2}, &  a > b,
            \\
            \sqrt{1-(a/b)^2}, &  a < b,
    \end{array}
    \right.
\end{equation}
it has been shown in Refs.~\cite{prd,prd2} that this possibility is achieved
if the eccentricities at decoupling is approximatively
\begin{equation}
\label{ecc-dec} e_{\rm dec}^2 \simeq \frac{\sqrt{15} \,
[ \, 3\sqrt{73} - 5 \, \mbox{sgn} (a-b) \, ]}{24} \, \frac{\mathcal{Q}_{\rm I}}{T_{\rm cmb}} \, ,
\end{equation}
where $\mbox{sgn} \, x$ is the sign function, 
$\mathcal{Q}_{\rm I}^2 \simeq 1200 \, \mu \mbox{K}^2$~\cite{WMAP7a} is the best-fit value
of the quadrupole amplitude for the $\Lambda$CDM concordance model, and
$T_{\rm cmb} \simeq 2.73$ K is the actual (average) CMB temperature.

Cosmic anisotropy could be triggered, as discussed in Refs.~\cite{prl,prd2},
by the presence in the Universe of a (almost) uniform magnetic field, or
topological defects such as a domain wall or a cosmic string. According to
Ref.~\cite{prd2}, instead, a possible cause of anisotropization could be Lorentz
symmetry violation during inflation which, in turn, generates magnetic fields
possessing planar symmetry at large cosmological scales. Recently, it has been
shown that a dark energy component with anisotropic equation of state has all the
requirements to give rise an ellipsoidal universe~\cite{ADE}.

Whatever is the source of anisotropy, its energy-momentum tensor has to be consistent
with planar symmetry:
\begin{equation}
\label{tensor1}
T^{\mu}_{\,\,\, \nu} = \mbox{diag} \, (\rho,-p_{\|},-p_{\|},-p_{\bot}),
\end{equation}
where $\rho$ is the energy density, while $p_{\|}$ and $p_{\bot}$ are
the ``longitudinal'' and ``normal'' pressures.

In the following, we consider a universe filled with a generic and unspecified
anisotropic component,
\begin{equation}
\label{tensor2}
{(T_A)}^{\mu}_{\;\; \nu} = \mbox{diag} \, (\rho^A,-p^A_{\|},-p^A_{\|},-p^A_{\bot}),
\end{equation}
which induces the planar symmetry and an isotropic contribution,
\begin{equation}
\label{tensor3} {(T_I)}^{\mu}_{\;\; \nu} = \mbox{diag} \, (\rho^I,-p^I, -p^I, -p^I),
\end{equation}
made up of three different components: a radiation component ($r$), a matter
component ($m$), and a cosmological constant component ($\Lambda$),
\begin{eqnarray}
\label{rhoI} && \rho^I = \rho_r + \rho_m + \rho_{\Lambda}, \\
\label{pI} && p^I = p_r + p_m + p_{\Lambda},
\end{eqnarray}
with equations of state: $p_{r} = \rho_{r}/3$, $p_{m} = 0$, and
$p_{\Lambda}=-\rho_{\Lambda}$.

Let us introduce, in the usual way, the cosmic shear, $\Sigma$,
and the ``mean Hubble parameter'', $H$, as
\begin{equation}
\label{ShearHubble} \Sigma \equiv (H_a - H)/H, \;\;\;
H \equiv \dot{A}/A,
\end{equation}
where $H_a \equiv \dot{a}/a$ and $A \equiv (a^2b)^{1/3}$ is the
``mean expansion parameter''.

Taking into account the above definitions, the
Einstein's equations for the cosmological model at hand read
\begin{eqnarray}
\label{E1} && \BS \bs \: (1-\Sigma^2) H^2 = \frac{8\pi G}{3} \, (\rho^I + \rho^A), \\
\label{E2} && \BS \bs \: (1-\Sigma + \Sigma^2) H^2 + [(2-\Sigma) H]^{\cdot} =
           -\frac{8\pi G}{3} \, (p^I + p_{\|}^A), \\
\label{E3} && \BS \bs \: (1+\Sigma)^2 H^2 + 2[(1+\Sigma) H]^{\cdot} =
           -\frac{8\pi G}{3} \, (p^I + p_{\bot}^A),
\end{eqnarray}
where a dot denotes a differentiation with respect to the cosmic time.

The source of anisotropy is proportional to the difference between the longitudinal and
normal pressures of the anisotropic fluid, as it is evident if we subtract side by side
Eqs.~(\ref{E2}) and (\ref{E3}):
\begin{equation}
\label{EvolShear} (H\Sigma)^{\cdot} + 3H^2\Sigma =
\frac{8\pi G}{3} \, (p_{\|}^A - p_{\bot}^A).
\end{equation}
For the sake of simplicity, we assume that all components are
noninteracting, so that they are separately conserved. For the
isotropic components we have $(T_X)^{\mu}_{\,\,\, \nu \, ;\mu} = 0$,
where $X=r,m,\Lambda$, which gives
\begin{eqnarray}
\label{rhoX}
&& \rho_r = \rho_r^{(0)} A^{-4}, \;\;\; \rho_m = \rho_m^{(0)} A^{-3},
\;\;\; \rho_\Lambda = \rho_\Lambda^{(0)},
\end{eqnarray}
where from now on an index ``0'' defines quantities evaluated at the actual time.
The conservation of the anisotropic part of the energy-momentum tensor,
$(T_A)^{\mu}_{\,\,\, \nu \, ;\mu} = 0$, gives instead
\begin{equation}
\label{EqEnergy} \dot{\rho}^A + [ \, 3(1+w) + 2 \delta \Sigma \,] \, H \rho^A = 0,
\end{equation}
where we have introduced the ``mean equation of state parameter'' $w$ and
the ``skewness'' $\delta$ as
\begin{equation}
\label{wdelta} w \equiv \frac{2p_{\|}^A + p_{\bot}^A}{3\rho^A},
\;\;\; \delta \equiv \frac{p_{\|}^A - p_{\bot}^A}{\rho^A}.
\end{equation}
Moreover, we assume that $w$ and $\delta$ are constants and that $\Sigma$ is a
small quantity (as we will verify {\it a posteriori}). Therefore, we can neglect
the second term in the square brackets of Eq.~(\ref{EqEnergy}), which simply gives:
\begin{equation}
\label{EqDE2} \rho_A = \rho_A^{(0)} A^{-3(1+w)}.
\end{equation}
Introducing the ``mean density parameters''
\begin{equation}
\label{OmegaX} \Omega_X \equiv \rho_X^{(0)}/\rho_c^{(0)}, \;\;\;
\rho_c^{(0)} \equiv \frac{3H_0^2}{8\pi G} \, ,
\end{equation}
where $X=r,m,\Lambda,A$, and taking into account Eqs.~(\ref{rhoX})-(\ref{OmegaX}),
the shear equation~(\ref{EvolShear}) can be solved to give
\begin{equation}
\label{Sigmat}  \Sigma(A) = \frac{\Sigma_0 + (E-E_0) \, \Omega_A \delta}{A^3H/H_0} \, ,
\end{equation}
where
\begin{equation}
\label{EqH}
H(A)/H_0 = \sqrt{\Omega_r A^{-4} + \Omega_m A^{-3} + \Omega_\Lambda + \Omega_A A^{-3(1+w)}}
\end{equation}
and we have defined the function
\begin{equation}
\label{xi} E(A) = \int_0^A \!\! \frac{dx}{x^{1+3w} H(x)/H_0} \, .
\end{equation}
Evaluating Eq.~(\ref{EqH}) at the present time gives
\begin{equation}
\label{Omegatot} \Omega_r + \Omega_m + \Omega_\Lambda + \Omega_A = 1,
\end{equation}
so that the density parameters are not all independent.

The standard (isotropic) $\Lambda$CDM concordance model fits, at a high level of accuracy,
cosmological astrophysical data coming from very disparate phenomena taking place in various
epoch of the Universe, such as~\cite{Weinberg} Big Bang Nucleosynthesis and Large Scale Structure
formation. For this reason, we only want to
consider anisotropic cosmological models which are very close to it. We then require
both that the energy density of the anisotropic component is negligibly small compared to
the energy densities of the isotropic ones (``subdominance condition''),
\begin{equation}
\label{Subdominance} \forall A \in [0,1]: \;\;
\rho_A(A) \ll \max_{A \in [0,1]} [\rho_r + \rho_m + \rho_\Lambda],
\end{equation}
%
and that the Universe isotropize at early times (``isotropization condition''),
\begin{equation}
\label{Isotropization} \lim_{A \rightarrow 0} \Sigma(A) = 0.
\end{equation}
Taking the limit $A \rightarrow 0$ in Eq.~(\ref{Sigmat}),
we find that the isotropization condition is satisfied if and only if
\begin{equation}
\label{Sigma0}  \Sigma_0 = E_0 \Omega_A \delta, \;\;\; w < 1/3.
\end{equation}
When $w=1/3$, condition~(\ref{Isotropization}) is never fulfilled. However, in this particular case,
we find that at early times $\Sigma$ approaches the constant value
\begin{equation}
\label{1/3}  \lim_{A \rightarrow 0} \Sigma(A) = \frac{\Omega_A \delta}{\Omega_r} \, , \;\;\; w = 1/3
\end{equation}
if Eq.~(\ref{Sigma0}) is satisfied. 
Even if the Universe
does not isotropize for $A \rightarrow 0$, it becomes almost isotropic for, as we will show,
the quantity $\Omega_A \delta$ is vanishingly small for $w=1/3$ (see middle-right panel of Fig.~1).


\begin{figure}[t]
\begin{center}
\includegraphics[clip,width=0.484\textwidth]{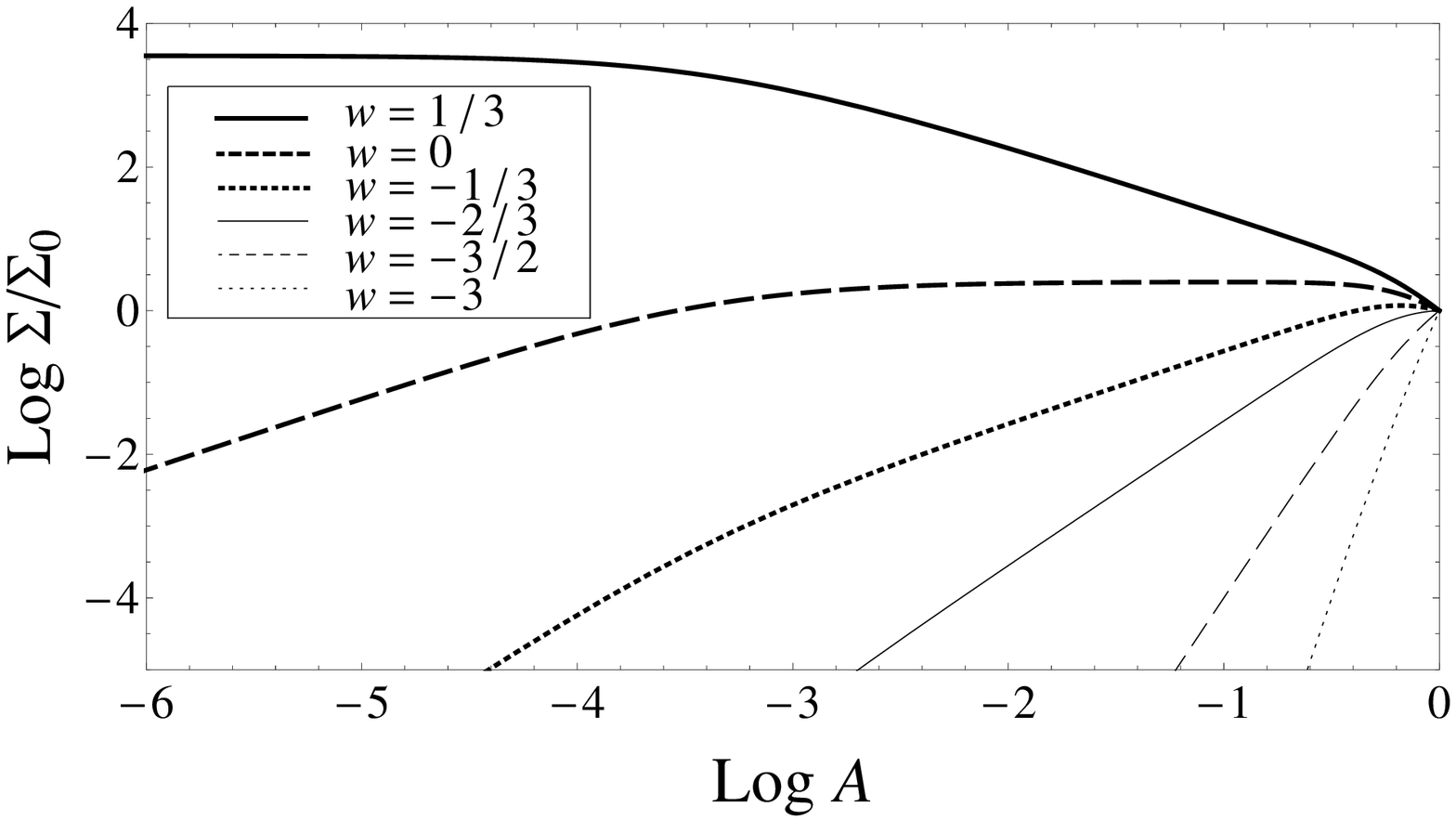}

\vspace*{0.3cm}

\includegraphics[clip,width=0.23\textwidth]{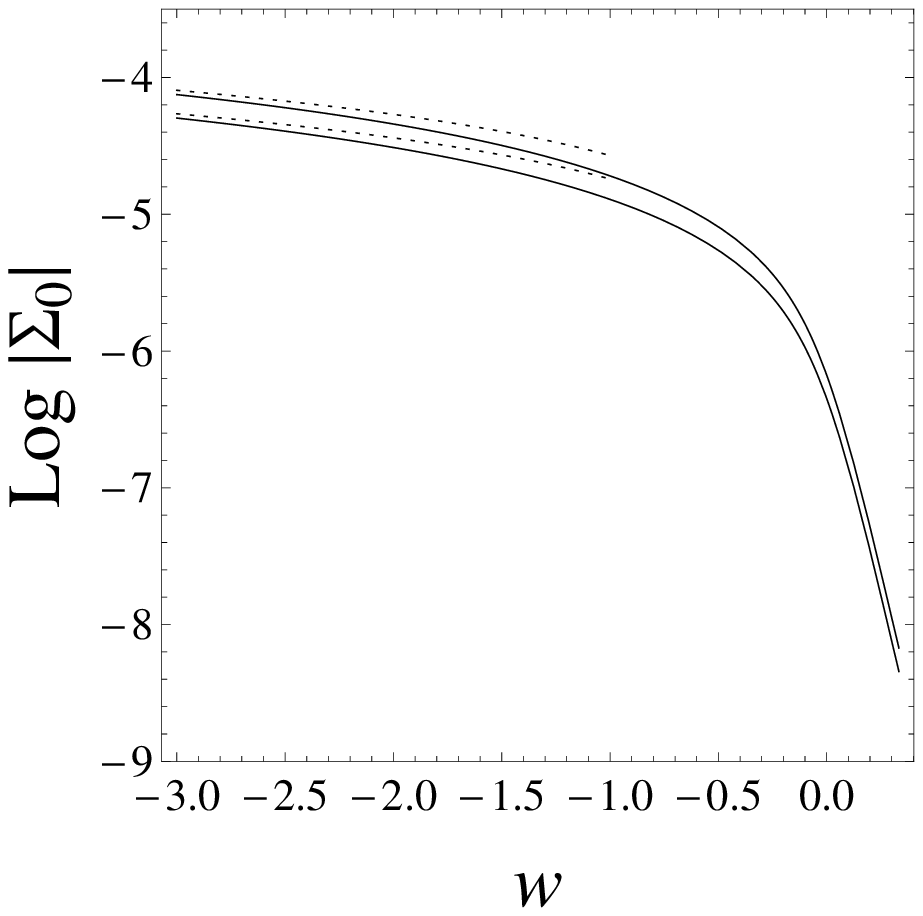}
\hspace*{0.2cm}
\includegraphics[clip,width=0.23\textwidth]{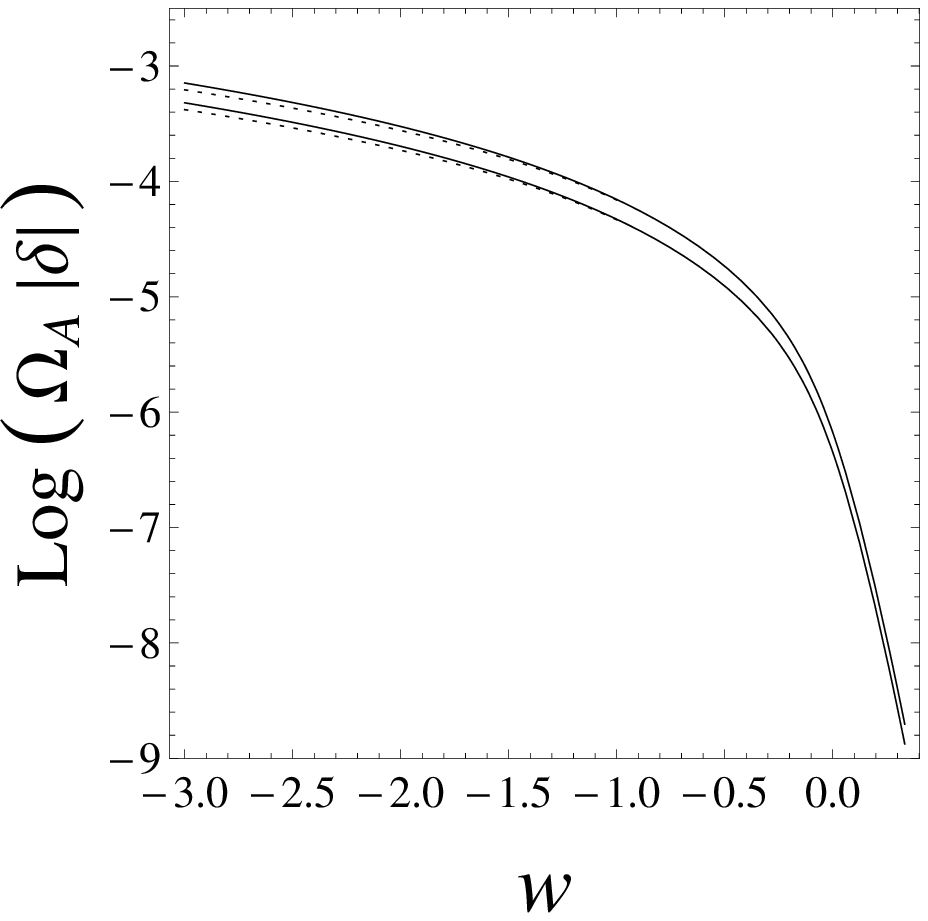}

\vspace*{0.3cm}

\includegraphics[clip,width=0.24\textwidth]{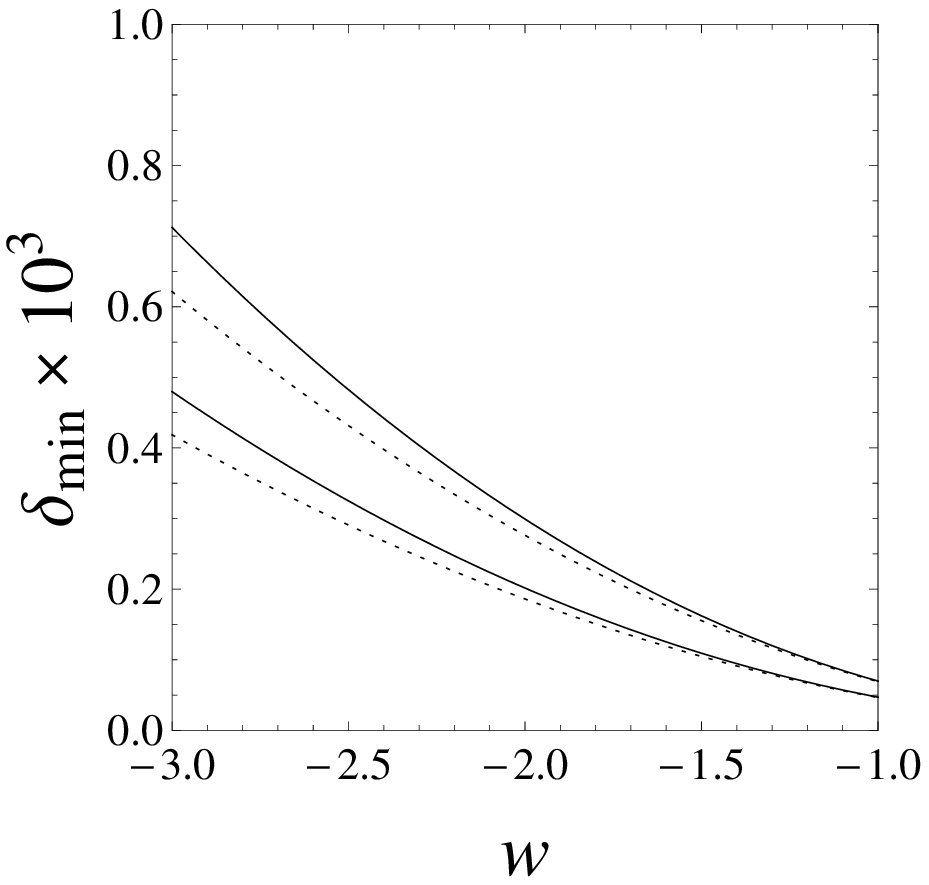}
\includegraphics[clip,width=0.23\textwidth]{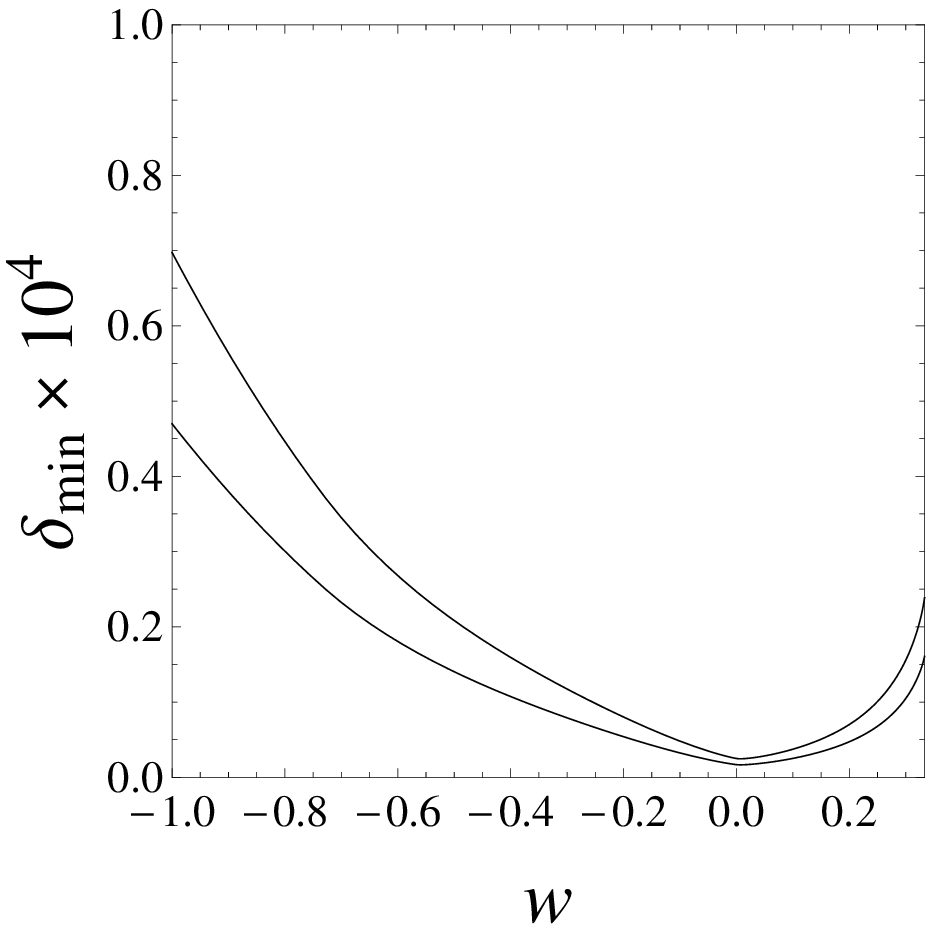}
\caption{{\it Upper panel.} The shear $\Sigma$ as a function of the mean expansion
parameter $A$ for different values of the mean equation of state parameter $w$.
{\it Middle panels.} The actual shear $\Sigma_0$ (left panel) and the mean density parameter
$\Omega_A$ multiplied by the skewness $\delta$ (right panel) as a function of $w$ for positive
(upper curves) and negative (lower curves) $\delta$, together with the asymptotic expansions
for $w \ll -1$ (dotted curves).
{\it Lower panels.} The minimum skewness $\delta_{\rm min}$ as a function of $w$
for $\delta > 0$ (upper curve) and $\delta < 0$ (lower curve), together with the
asymptotic expansions for $w \ll -1$ (dotted curves).}
\end{center}
\end{figure}


Assuming that all the above conditions are fulfilled, we can write the shear as
\begin{equation}
\label{Sigmat2}  \Sigma(A) = \Sigma_0 \: \frac{E/E_0}{A^3H/H_0} \, ,
\end{equation}
together with its asymptotic expansion for $A \rightarrow 0$:
\begin{equation}
\label{Sigmat0}  A \ll 1: \;\;
\Sigma(A) \simeq \frac{\delta}{2-3w} \, \frac{\Omega_A}{\Omega_r} \, A^{1-3w}.
\end{equation}
In Fig.~1 (upper panel), we show the shear $\Sigma$ as a function of the mean expansion
parameter $A$.
The smallness of shear (which we will verify in a moment) assures that the values of
mean parameters $\Omega_r$ and $\Omega_\Lambda$ are very close to the analogous ones
for the isotropic standard cosmological model. For this reason, we take
in this paper:
$\Omega_r \simeq \Omega_r^{(\rm isotropic)} \simeq 0.83 \times 10^{-4}$~\cite{Kolb}
and $\Omega_\Lambda \simeq \Omega_\Lambda^{(\rm isotropic)} \simeq 0.73$~\cite{WMAP7b}.

We want now to connect the density parameter $\Omega_A$ to the eccentricity at decoupling,
since the anisotropic component is supposed to be the cause of the anisotropization of the
Universe at decoupling. To this end, we observe that the eccentricity and the shear are
connected by the following relation:
\begin{equation}
\label{EccSigma} e^2 = 6 \, \mbox{sgn} (a-b)
\int_1^A \frac{dx}{x} \, \Sigma(x),
\end{equation}
valid for small eccentricities, $e \ll 1$, and coming from definitions~(\ref{ecc-def})
and (\ref{ShearHubble}). By evaluating the above equation at the time of decoupling,
and using Eqs.~(\ref{Sigma0}) and (\ref{Sigmat2}), we get
\begin{equation}
\label{sgn} \mbox{sgn} (a-b) = -\mbox{sgn} \, \delta
\end{equation}
together with
\begin{equation}
\label{Omega-ecc} \Omega_A = \frac{c_\Omega(w)}{|\delta|} \: e_{\rm dec}^2,
\end{equation}
and we recast $\Sigma_0$ in the form
\begin{equation}
\label{Sigma0-ecc} |\Sigma_0| = c_\Sigma(w) \, e_{\rm dec}^2,
\end{equation}
where
\begin{eqnarray}
\label{cOmega} c_\Omega \!\!& \equiv &\!\!
\left[6 \! \int_{A_{\rm dec}}^1 \frac{dx}{x^4} \, \frac{E(x)}{H(x)/H_0} \right]^{\!-1} \!,
\\
\label{cSigma0} c_\Sigma \!\!& \equiv &\!\! E_0 c_\Omega.
\end{eqnarray}
Here, $A_{\rm dec} = A(t_{\rm dec})$ is the mean expansion parameter evaluated at the
time of decoupling and will be simply taken to be $A_{\rm dec} = 1/(1+z_{\rm dec})$,
with $z_{\rm dec} \simeq 1090$~\cite{WMAP7b} the redshift at decoupling.

In Fig.~1 (middle panels), we show the actual shear $\Sigma_0$ and the density parameter
$\Omega_A$ as a function of the mean equation of state parameter $w$, with $e_{\rm dec}$
given by Eq.~(\ref{ecc-dec}), together with the asymptotic expansions for large negative
values of $w$:
\begin{eqnarray}
\label{Sigma0ApproxInfty}
&& w \ll -1: \;\;
|\Sigma_0| \simeq \frac{1}{2} \sqrt{\Omega_\Lambda} \, e_{\rm dec}^2 \, |w|, \\
\label{OmegaApproxInfty}
&& w \ll -1: \;\;
\Omega_A \simeq \frac{3}{2} \, \Omega_\Lambda e_{\rm dec}^2 \, \frac{w^2}{|\delta|} \, .
\end{eqnarray}
As it is clear from the upper panel and the left middle panel of Fig.~1,
the cosmic shear is always much smaller than unity, as we have previously assumed.
Moreover, the actual fraction of energy associated to the anisotropic component
(see the right middle panel of Fig.~1) is negligible with respect to those of
dark matter and cosmological constant if $\delta$ is not too small in absolute value.
Very small values of $|\delta|$ are not allowed, however, by the
subdominance condition~(\ref{Subdominance}), which taking into account
Eq.~(\ref{Omega-ecc}) can be rewritten as
\begin{equation}
\label{deltamin} |\delta| \gg \delta_{\rm min} = c_\delta(w) \, e_{\rm dec}^2,
\end{equation}
where
\begin{equation}
\label{cdelta} c_\delta \equiv \frac{c_\Omega}{\max_{A \in [0,1]}[A^{3(1+w)} H^2(A)/H_0^2]} \, .
\end{equation}
In Fig.~1 (lower panels), we plot the minimum skewness $\delta_{\rm min}$
as a function of $w$, together with the asymptotic expansion:
\begin{equation}
\label{dwApproxInfty}
w \ll -1: \;\; \delta_{\rm min} \simeq \frac{3}{2} \, \Omega_\Lambda e_{\rm dec}^2 \, w^2.
\end{equation}
Finally, in Tab.~1, we show the results for some particular and physically relevant case
of anisotropic component, such as planar magnetic field (discussed in Ref.~\cite{prd2}),
uniform magnetic field, cosmic string, and domain wall.


\begin{table}
\begin{center}
\caption{Some components with anisotropic equation of state which could give rise to an
ellipsoidal universe: The skewness $\delta$ is the deviation from isotropy of the equation
of state, whose mean parameter is $w$; the mean density parameter is $\Omega_A$, while the
actual amount of anisotropy in the geometry of the universe is the shear $\Sigma_0$.}
\vspace{0.5cm}
\begin{tabular}{lllllll}

\hline \hline

&Component     &~~~~~\,$w$   &~~~~$\delta$  &~~~~~\;\,$\Omega_A$      &~~~~\;\;$\Sigma_0$      \\
\hline
&$B$ (planar)  &~~~~\,$1/3$  &~\:$-1$       &~~$1.3 \times 10^{-9}$   &~\:$4.5 \times 10^{-9}$ \\
\hline
&$B$ (uniform) &~~~~\,$1/3$  &~~~\;$2$      &~~$9.9 \times 10^{-10}$  &~\:$6.7 \times 10^{-9}$ \\
\hline
&String        &~~$-1/3$     &~~~\;$1$      &~~$9.2 \times 10^{-6}$   &~\:$5.0 \times 10^{-6}$ \\
\hline
&Domain wall   &~~$-2/3$     &~\:$-1$       &~~$2.1 \times 10^{-5}$   &~\:$7.7 \times 10^{-6}$ \\

\hline \hline

\end{tabular}
\end{center}
\end{table}


Looking at the Table and again at Fig.~1, we get that (realistic) values of $|w|$ of order unity
give at most $\Sigma_0$ of order $10^{-4}$. This result will be important for the following
discussion about the possibility to detect cosmic anisotropy by cosmic parallax effects.


\section{\normalsize{III. Cosmic Parallax and a Preferred Direction in the Sky}}
\renewcommand{\thesection}{\arabic{section}}

The angle $\gamma$ between two sources in an ellipsoidal universe,
as view by an observer centered at the origin of reference system,
is given by~\cite{Trodden}:
\begin{equation}
\label{parallax1} \cos\gamma(t) = \frac{\sum_i a_i^{-2}\hat{p}_i\hat{q}_i}
{\left( \sum_i a_i^{-2} \hat{p}_i^2 \right)^{\!1/2}
\left( \sum_i a_i^{-2} \hat{q}_i^2 \right)^{\!1/2}} \, ,
\end{equation}
where $a_1 \equiv a_2 \equiv a$ and $a_3 \equiv b$ are the expansion parameters,
while $\hat{p}$ and $\hat{q}$ are the direction cosines defining the angular position
of the sources. If the Universe expands anisotropically, then the angle $\gamma$
is time dependent and one can hope to detect its temporal variation, the so-called
{\it cosmic parallax} $\Delta\gamma$, looking at two different sources at two different
times.

It is worth stressing that the above result applies to the case
where the axis of symmetry is directed along the $z$-axis. We may, however,
easily generalize this result to the case where the symmetry axis
is directed along a particular direction $(b,l) = (b_A,l_A)$ in the
galactic coordinate system, where $b$ and $l$ are, respectively,
the galactic latitude and galactic longitude.
To this end, we perform a rotation
\begin{equation}
\label{Rotation} \mathcal{R} \equiv \mathcal{R}_{x}(\pi/2-b_A) \,
\mathcal{R}_{z}(\pi/2+l_A)
\end{equation}
of the coordinate system $(x,y,z)$, where $\mathcal{R}_{x}(\pi/2-b_A)$ and
$\mathcal{R}_{z}(\pi/2+l_A)$ are rotations of angles $\pi/2-b_A$ and
$\pi/2+l_A$ about the $x$- and $z$-axis, respectively.
In the galactic coordinate system the axis of symmetry is defined by the
direction cosines
\begin{equation}
\label{nA} \hat{n}_A \equiv (\cos\!b_A \cos l_A, \cos\!b_A \sin l_A, \sin\!b_A),
\end{equation}
while $\hat{p}$ and $\hat{q}$ are defined by the new
direction cosines $\hat{n} = \mathcal{R}^{-1} \hat{p}$ and
$\hat{n}' = \mathcal{R}^{-1} \hat{q}$, where
\begin{eqnarray}
\label{n} && \hat{n} \equiv (\cos\!b \cos l, \cos\!b \sin l, \sin\!b), \\
\label{n} && \hat{n}' \equiv (\cos\!b' \cos l', \cos\!b' \sin l', \sin\!b').
\end{eqnarray}
In the galactic coordinate system, with the axis of symmetry defined by the
unit vector $\hat{n}_A$, the angle $\gamma$ is
\begin{equation}
\label{parallax2} \cos\gamma(t) = \frac{\cos\gamma_0 - e^2(\cos\gamma_0 - \cos\alpha \cos\alpha')}
{\left( 1-e^2 \sin^2\alpha \right)^{1/2} \left( 1-e^2 \sin^2\alpha' \right)^{1/2}} \,
\end{equation}
when $a > b$, or
\begin{equation}
\label{parallax3} \cos\gamma(t) = \frac{\cos\gamma_0 - e^2 \cos\alpha \cos\alpha'}
{\left( 1-e^2 \cos^2\alpha \right)^{1/2} \left( 1-e^2 \cos^2\alpha' \right)^{1/2}} \, ,
\end{equation}
if $a < b$, where we have introduced the eccentricity $e$.
The quantities $\gamma_0$, $\alpha$ and $\alpha'$ are the angles between,
respectively, $\hat{n}$ and $\hat{n}'$, $\hat{n}$ and $\hat{n}_A$, and $\hat{n}'$ and $\hat{n}_A$:
\begin{eqnarray}
\label{angles} \bs \BS && \cos\gamma_0 \equiv \hat{n} \cdot \hat{n}'
               = \sin\!b \sin\!b' + \cos\!b \cos\!b' \cos(l-l'),    \nonumber \\
               \bs \BS && \cos\alpha \equiv \hat{n} \cdot \hat{n}_A
               = \sin\!b \sin\!b_A + \cos\!b \cos\!b_A \cos(l-l_A), \nonumber \\
               \bs \BS && \cos\alpha' \equiv \hat{n}' \! \cdot \hat{n}_A
               = \sin\!b' \! \sin\!b_A + \cos\!b' \! \cos\!b_A \cos(l'\!-l_A). \nonumber \\
\end{eqnarray}
Differentiating Eqs.~(\ref{parallax2}) and (\ref{parallax3}), we straightforwardly obtain
the expression for the cosmic parallax $\Delta\gamma$ in a small time interval $\Delta t$
centered around $t_0$ and caused by an anisotropic expansion of the Universe:
\begin{equation}
\label{parallax4} \Delta\gamma \simeq \left.\frac{d\gamma}{dt}\right|_{t=t_0} \! \Delta t
\; = \; 3 \Upsilon H_0 \Sigma_0 \Delta t,
\end{equation}
where we introduced the ``modulating function'' $\Upsilon$ as
\begin{equation}
\label{Upsilon} 
\Upsilon(b_A,l_A;b,l;b',l') \equiv
\cot\!\gamma_0 (\cos^2\alpha + \cos^2\alpha')
- 2\csc\!\gamma_0\cos\alpha \cos\alpha',
\end{equation}
whose values are in the interval $[-1,1]$.

In the case where the axis of symmetry is directed along the $z$-axis
($b_A = \pi/2$), one easily recovers the results of Ref.~\cite{Quercellini2a,Quercellini2b}.

Equation~(\ref{Upsilon}) is an involved expression of the coordinates of the axis of
symmetry $(b_A,l_A)$ and of the directions of the two sources $(b,l)$ and $(b',l')$.


\begin{figure}
\begin{center}
\includegraphics[clip,width=0.484\textwidth]{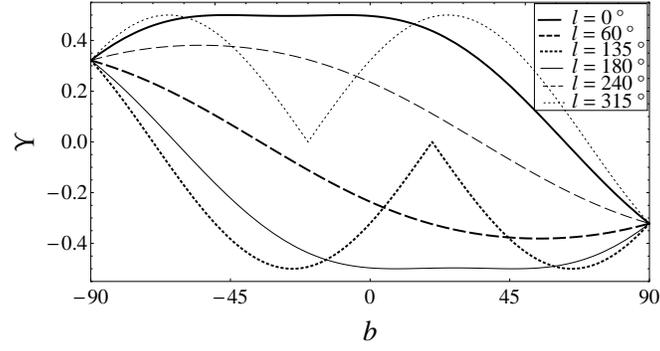}
\caption{The modulating function $\Upsilon$ [see Eqs.~(\ref{angles})-(\ref{Upsilon})]
versus $b$ for $(b',l') = (b_A,l_A) = (b_{AE},l_{AE}) = (20^\circ,135^\circ)$ for
different values of $l$. The thick dotted curve is Eq.~(\ref{Upsilon2}).}
\end{center}
\end{figure}


There are, however, some interesting cases where it gives very simple results.
For example, if one of the two sources is in the direction of the symmetry axis,
let us say $\hat{n}' \equiv \hat{n}_A$, 
and the other one has the same longitude of the symmetry axis, $l=l_A$,
we simply have 
\begin{equation}
\label{Upsilon2}
\Upsilon(b_A,l_A;b,l_A;b_A,l_A) = \frac{1}{2} \sin\!2(b-b_A).
\end{equation}
In Fig.~2, we plot the modulating function versus the latitude $b$ of one source for different
values of its longitude $l$. We take the other source pointing in the direction of the symmetry
axis which we assume to be the axis of evil~\cite{Groeneboom}:
\begin{equation}
\label{AE}
(b_{AE},l_{AE}) \simeq (20^\circ,135^\circ).
\end{equation}
In principle, cosmic parallax data from planned astronomic missions, such as Gaia,
could allow, other than constraining the cosmic shear [see Eq.~(\ref{parallax4})], to determine
the direction of the symmetry axis {\it via} the analysis of the modulating function. Indeed,
the latter is very sensitive to the position of the axis since, for example, if we look at
two sources, one along the $z$-axis and the other along the $x$-axis [$(b',l')=(0,0)$], we find:
\begin{equation}
\label{Upsilon3}
\Upsilon(b_A,l_A;\pi/2,l;0,0) = \sin 2b_A \cos l_A.
\end{equation}
The modulating function, in this case, is zero if the symmetry axis coincides with the
$z$-axis (as supposed in Refs.~\cite{Quercellini2a,Quercellini2b,Trodden}) or, in general, when lies in
the galactic plane ($b_A=0$) or in a plane perpendicular to it ($b_A = \pm \pi/2$),
maximal ($\Upsilon = 1$) for $(b_A,l_A) = (45^\circ,0^\circ)$ or $(-45^\circ,180^\circ)$,
and minimal ($\Upsilon = -1$) in the directions $(b_A,l_A) = (45^\circ,180^\circ)$ and
$(-45^\circ,0^\circ)$.

However, if one takes into account that the maximum precision of Gaia in measuring cosmic
parallax is about $6\mu\as$~\cite{Sozzetti}, a large cosmic anisotropy of order
$|\Sigma_0| \sim 10^{-2}$ is required in order to extract information from the modulating
function. In fact, assuming a capability to detect the angular position of two sources at two
different times separated by $\Delta t = 10 \yr$, we can conveniently rewrite
Eq.~(\ref{parallax4}) as
\begin{equation}
\label{parallax5} \Delta\gamma \simeq 4.4 \, \Upsilon \:
\frac{h}{0.70} \: \frac{\Sigma_0}{10^{-2}} \: \frac{\Delta t}{10 \yr} \: \mu\as,
\end{equation}
where the little-$h$ constant, $H_0 \equiv 100 \, h \,\km / \s / \Mpc$,
is about $h^{(\rm isotropic)} \simeq 0.70$~\cite{WMAP7b} in the isotropic
standard cosmological model.

The ellipsoidal universe proposal, allowing at maximum cosmic shears of order
$|\Sigma_0| \sim 10^{-4}$ (see Fig.~1), is then not testable, for the time being,
by cosmic parallax measurements.


\section{\normalsize{IV. Conclusions}}
\renewcommand{\thesection}{\arabic{section}}

Homogeneity and isotropy of the Universe on large cosmological scales are the foundation of
standard cosmology. Although they are simply {\it assumed} as the basis of observational and
theoretical cosmology, experimental data seem, up today, to confirm and support the grounding
of those hypotheses. In particular, two decades of observations of the cosmic microwave
background radiation and analysis of magnitude-redshift data on type Ia supernovae, have firmly
established that, if there are deviations from isotropy, they must be very tiny.

If on the one hand the predictions of the isotropic $\Lambda$CDM concordance model are in
streaking agreement with a huge amount of observational data collected in the last years,
on the other hand some tension, between theoretical and inferred value of the CMB quadrupole
anisotropy, still persists in CMB data, even after the recent results of seven years WMAP
observations.

The deficit of power on large angular scales, emerged since the first data of COBE back in 1992
and now known as ``quadrupole problem'', could be a hint of anisotropization of the Universe
at cosmic scales. Indeed, the simplest anisotropic cosmological model, namely a Bianchi type I
characterized by a plane-symmetric line element, has been shown to be compatible with all
CMB and supernovae data so far gathered and analyzed and, moreover, allow for a solution to the
quadrupole problem.

In this paper, we have further investigated such a kind of nonstandard cosmological model, named
``ellipsoidal universe'', turning our attention to a possible signature that could be detected
in high-precision astrometric observations, that is a ``cosmic parallax'' effect. As already noticed
in the literature, the relative angular position of two distant sources changes in time if the
Universe expands anisotropically. Intuitively to understand but difficult to measure, this effect
of parallax represents a unique possibility to directly test the hypothesis of isotropy/anisotropy
of the Universe.

We have shown, indeed, that planned astrometric missions such as Gaia, have too low accuracies both
to appreciate the small amount of anisotropy and to confirm the existence of a preferred direction
in the sky predicted by the ellipsoidal universe model.

This, however, does not exclude the possibility that, in a not-too-far future, Gaia-like missions
with enhanced sensitivity of about at least two order of magnitude could either see a signal of
cosmic anisotropy or completely rule out the ellipsoidal universe model.


\end{document}